\documentclass[10pt,twocolumn,letterpaper]{article}

\usepackage[pagenumbers]{wacv} 

\usepackage{graphicx}
\usepackage{epstopdf}
\usepackage{amsmath}
\usepackage{amssymb}
\usepackage{booktabs}
\usepackage{lipsum} 

\usepackage{colortbl}
\usepackage{xcolor}
\usepackage{cite}
\usepackage{float}
\usepackage{tabularray,varwidth,enumitem}
\usepackage{array}
\usepackage{multicol}
\usepackage{multirow, makecell}
\usepackage[parfill]{parskip}
\usepackage{algorithm}
\usepackage{algorithmic}
\usepackage{hhline}
\usepackage{caption}
\usepackage{pifont}

\newcommand{\cmark}{\checkmark} 
\newcommand{\xmark}{\ding{55}}  

\newcommand{\second}[1]{\textbf{\textcolor{blue}{#1}}}
\newcommand{\best}[1]{\textbf{\textcolor{red}{#1}}}
\newcommand{\figref}[1]{Figure \ref{#1}}

\usepackage[pagebackref,breaklinks,colorlinks]{hyperref}

\usepackage[capitalize]{cleveref}
\crefname{section}{Sec.}{Secs.}
\Crefname{section}{Section}{Sections}
\Crefname{table}{Table}{Tables}
\crefname{table}{Tab.}{Tabs.}

\def\wacvPaperID{2208} 
\def\confName{WACV}
\def\confYear{2025}

\begin{document}


\title{CT to PET Translation: \\ A Large-scale Dataset and Domain-Knowledge-Guided Diffusion Approach}



\author{
Dac Thai Nguyen$^{1}$, Trung Thanh Nguyen$^{2}$, Huu Tien Nguyen$^{1}$, Thanh Trung Nguyen$^{3}$, \\
Huy Hieu Pham$^{4}$, Thanh Hung Nguyen$^{1}$, Thao Nguyen Truong$^{5}$, and Phi Le Nguyen$^{1}$\\
$^{1}$Hanoi University of Science and Technology, Vietnam;
$^{2}$Nagoya Univeristy, Japan \\ $^{3}$108 Military Central Hospital, Vietnam; $^{4}$VinUniversity, Vietnam\\
$^{5}$National Institute of Advanced Industrial Science and Technology, Japan\\
}
\maketitle

\begin{abstract}
Positron Emission Tomography (PET) and Computed Tomography (CT) are essential for diagnosing, staging, and monitoring various diseases, particularly cancer. Despite their importance, the use of PET/CT systems is limited by the necessity for radioactive materials, the scarcity of PET scanners, and the high cost associated with PET imaging. In contrast, CT scanners are more widely available and significantly less expensive. In response to these challenges, our study addresses the issue of generating PET images from CT images, aiming to reduce both the medical examination cost and the associated health risks for patients. Our contributions are twofold: First, we introduce a conditional diffusion model named \textbf{CPDM}, which, to our knowledge, is one of the initial attempts to employ a diffusion model for translating from CT to PET images. Second, we provide the largest CT-PET dataset to date, comprising 2,028,628 paired CT-PET images, which facilitates the training and evaluation of CT-to-PET translation models. For the CPDM model, we incorporate domain knowledge to develop two conditional maps: the Attention map and the Attenuation map. The former helps the diffusion process focus on areas of interest, while the latter improves PET data correction and ensures accurate diagnostic information. Experimental evaluations across various benchmarks demonstrate that CPDM surpasses existing methods in generating high-quality PET images in terms of multiple metrics. The source code and data samples are available at \url{https://github.com/thanhhff/CPDM}. 
\end{abstract}

\section{Introduction}
\label{sec:introduction}
The combination of PET and CT has markedly enhanced oncology imaging by integrating PET's functional imaging capabilities with the anatomical detail of CT scans. This integration boosts accuracy in diagnosis, staging, and evaluation of cancer treatments ~\cite{wei2021artificial, vijayakumar2022changing}. 
However, the application of PET in medical examination and treatment faces several key challenges: (1) PET imaging requires the injection of a small amount of radioactive material, which may pose health risks\cite{Ben_Cohen_2017}; (2) PET scanners are limited in availability, particularly in underdeveloped regions \cite{cherry2018total, alameen2021radiobiological}; and (3) the high cost of PET imaging imposes a financial burden on patients\cite{Ben_Cohen_2017}.
In contrast, CT imaging mitigates many of these issues, primarily because of its lower cost, typically around half that of PET imaging, thereby reducing the financial burden on patients.
Therefore, leveraging artificial intelligence to generate PET images from CT scans offers a viable solution to address the aforementioned challenges. This approach also holds substantial practical significance and high applicability for patient care.

\begin{figure}[t]
    \centering
    \includegraphics[width=1.0\linewidth]{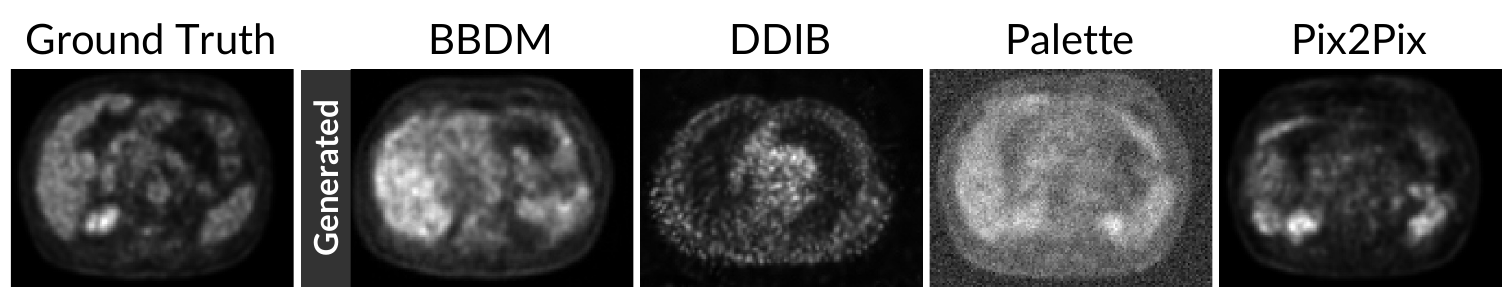}
    \caption{\textbf{Examples of synthetic PET images produced by natural image generative models (BBDM~\cite{li2023bbdm}, DDIB~\cite{su2022dual}, Palette~\cite{saharia2022palette}, and Pix2Pix~\cite{wang2018highresolutionimagesynthesissemantic}).} The generated images do not accurately replicate the ground truth. }
    \label{fig:i2i_natural}
    \vspace{-12pt}
\end{figure}
The challenge of converting CT to PET images falls within the domain of Image-to-Image (I2I) translation, a prominent area of research in computer vision. Numerous methods have been proposed to address this task, with generative models such as Generative Adversarial Network (GAN)~\cite{goodfellow2014generative} emerging as the leading approach. Based on GAN, Pix2Pix~\cite{isola2017image} made significant strides in I2I translation tasks with its model consisting of a generator and a discriminator. Since then, numerous GAN variants, such as CycleGAN~\cite{zhu2017unpaired}, UVCGAN~\cite{torbunov2023uvcgan}, and SelectionGAN~\cite{tang2019multi}, have been introduced. 
Despite their ability to generate synthetic images that closely resemble real ones, GAN-based models are often hindered by training instability and sensitivity to hyperparameters. Recently, diffusion model~\cite{croitoru2023diffusion} has emerged as a promising alternative, noted for their stability during training and ability to generate highly realistic images~\cite{dhariwal2021diffusion}.
Diffusion models have been applied to address the I2I translation problem in several studies~\cite{saharia2022palette, meng2021sdedit, li2023bbdm}. However, as shown in Figure~\ref{fig:i2i_natural}, their research primarily focuses on natural images, leading to suboptimal performance when applied to medical images, particularly in the CT-to-PET translation task.
Several studies have concentrated on addressing the I2I translation problem for medical images, such as those in~\cite{kong2021breaking, armanious2020medgan, liang2024leveraging, zhu2020cross}.  
However, current research encounters certain challenges, including: (1) A considerable amount of current I2I translation research in medical imaging
 is dedicated to transforming between two akin domains, such as MRI T1 to MRI T2~\cite{kong2021breaking, yang2020mri, wang2023spatial}, CBCT to CT~\cite{liang2024leveraging}. There is a noticeable shortage of research tackling translation between entirely distinct modalities, especially, from CT images to PET; (2) The dominant approach based on generative models entails generating a random image that conforms to the distribution of the target domain. While these models utilize the source image to guide the generation process, thereby enhancing the resemblance of the generated image to the ground truth, they fail to mitigate the inherent stochasticity of the generated images; (3) While the natural I2I translation benefits from large datasets \cite{li2023bbdm}, the domain of medical I2I, and CT-to-PET in particular, remains significantly data-constrained. This scarcity of data has hindered advancement in these fields.
 
To tackle the aforementioned challenges, this study focuses on a significant yet underexplored issue: CT-to-PET translation. Alongside providing an extensive dataset, we aim to develop a CT-to-PET translation model with the following objectives:
\begin{itemize}
    \item \textbf{Enhancing the deterministic nature of the generated PET image:} Rather than relying on conventional generative models that aim to generate random images, we employ a diffusion model directly mapping from the source domain to the target domain. In essence, this diffusion model learns a process to directly convert a CT image into a corresponding PET image.
    \item \textbf{Improving the accuracy of the generated PET image:} We integrate domain-specific knowledge to guide the diffusion process. Specifically, we utilize two conditional maps, namely the \textit{Attention map} and the \textit{Attenuation map}, to steer the diffusion process. The former essentially highlights pixels that are likely to be within the PET region of interest. Meanwhile, the latter, vital for accurate PET image quantification, computes photon attenuation at 511 keV from the CT scan, improving PET data correction and securing precise diagnostic information.
\end{itemize}

The main contributions of this paper include:
\begin{itemize}
    \item We propose \textbf{CPDM} (\textbf{C}T-to-\textbf{P}ET \textbf{D}iffusion \textbf{M}odel), a novel approach that leverages a Brownian Bridge process-based diffusion model to directly synthesize PET images from CT scans. 
    To refine the quality of the generated PET images, we guide the diffusion process with the domain knowledge from our additional conditional maps, i.e., Attention and Attenuation maps.
    \item We offer a comprehensive large-scale PET/CT medical dataset comprising 2,028,628 paired CT-PET images. To our knowledge, this represents the largest CT-PET dataset available. This dataset is curated to support developing and evaluating advanced medical image translation methods, providing a valuable resource for the research community.
    \item We conduct comprehensive experiments to evaluate the performance of CPDM against state-of-the-art (SOTA) image translation methods. The results demonstrate that CPDM outperforms SOTA methods across multiple metrics for PET image quality.
\end{itemize}



\section{Related Work}
\label{sec:related work}
In this section, we review existing CT and PET datasets in Section~\ref{subsec:dataset}, followed by a discussion of recent research on I2I translation in Section~\ref{subsec:midecial_image_translation}.

\subsection{CT and PET Datasets}
\label{subsec:dataset}
The advancement of cancer imaging research has been significantly enhanced by the availability of large-scale 3D PET/CT datasets. 
The RIDER Lung PET-CT\cite{muzi2015riderlungpetct} (2015) and Lung-PET-CT-Dx\cite{li2020lungpetctdx} (2020) datasets, both focus on lung cancer, provide PET and CT scans that have been instrumental in developing multimodal imaging techniques and machine learning models for lung cancer diagnosis.
Expanding the anatomical scope, the Head-Neck-PET-CT\cite{vallieres2017headneckpetct} (2017) dataset offers specialized imaging data for head and neck cancers, while the FDG-PET-CT-Lesions\cite{gatidis2022fdgpetctlesions} (2022) dataset includes comprehensive whole-body PET/CT scans with manual annotations performed using specialized software. 
These datasets are pivotal in advancing research on lesion detection and comprehensive cancer diagnostics across different body regions.

Despite the significant contributions of these datasets, supervised learning tasks such as I2I translation require precisely paired PET and CT images on a slice-by-slice basis, a criterion not typically met by existing datasets. 
Furthermore, generative tasks, including image generation and translation, generally require extensive datasets to develop high-quality models, a need that is not adequately fulfilled by current CT-PET datasets (refer to Table~\ref{tab:overview_ctpet_image_datasets}). 
In this context, our study contributes a large-scale paired CT-PET dataset, which is intended not only to advance the CT2PET translation problem but also to support research involving CT, PET, and medical imaging in general.
 
\begin{table}[t]
\centering
\small
\caption{Overview of our dataset and existing CT and PET datasets (sorted by their published year). \textbf{Auto-Paired} means that corresponding CT and PET slices are automatically paired slice-by-slice, a feature missing in existing datasets, which only provide unpaired image series in DICOM format.}
\label{tab:overview_ctpet_image_datasets}
\setlength\tabcolsep{3pt} 
\resizebox{\linewidth}{!}
{
    \begin{tabular}{l|c|l|r|r}
    \toprule
    \textbf{Name of Dataset} & 
    \begin{tabular}[c]{@{}l@{}}\textbf{Auto}\\\textbf{Paired}\end{tabular} &
    \textbf{Body Parts} & 
    \begin{tabular}[c]{@{}l@{}}\textbf{Total} \\ \textbf{Slices}\end{tabular}& 
    \begin{tabular}[c]{@{}r@{}}\textbf{\# of}\\ \textbf{Studies} \end{tabular}
\\ 
\midrule
RIDER Lung PET-CT~\cite{muzi2015riderlungpetct} & \xmark & Lung & 266K & 274 \\  
Head-Neck-PET-CT~\cite{vallieres2017headneckpetct} & \xmark & Head \& Neck & 123K & 504 \\  
Lung-PET-CT-Dx~\cite{li2020lungpetctdx} & \xmark & Lung & 251K & 430 \\ 
FDG-PET-CT-Lesions~\cite{gatidis2022fdgpetctlesions} & \xmark & Whole body & 917K & 1,014 \\ 
\midrule
\textbf{Our Dataset} & \cmark & Whole body & 2M & 3,454 \\ 
\bottomrule
\end{tabular}
}
\vspace{-0.1cm}
\end{table}
\subsection{Image-to-Image Translation}
\label{subsec:midecial_image_translation}
\noindent \textbf{Generative Models for Natural Images.}
Generative models have been crucial in advancing I2I translation tasks, especially for natural images. Pix2Pix~\cite{wang2018highresolutionimagesynthesissemantic} pioneers conditional GANs to translate images from one domain to another with controlled output generation. While this model demonstrates significant progress, it struggles to produce diverse outputs due to its one-to-one mapping approach. 
Subsequent models, CycleGAN~\cite{zhu2017unpaired} and DRIT++~\cite{lee2019dritdiverseimagetoimagetranslation} seek to address these limitations by enabling unpaired image translation and generating diverse samples. 
However, these models face challenges in training stability and mode collapse, restricting their performance in complex tasks.
More recently, diffusion-based models have emerged as a promising alternative for image synthesis, offering increased stability and high-quality outputs. 
Palette~\cite{saharia2022palette} and SDEdit~\cite{meng2021sdedit} demonstrate superior performance across various I2I tasks without requiring task-specific tuning, making them flexible solutions across domains. 
Subsequently, LDM~\cite{rombach2022high} improves efficiency by conducting the diffusion process in the latent space of pre-trained models. 
Building on these advancements, BBDM~\cite{li2023bbdm} introduces a novel Brownian Bridge diffusion process for direct image domain translation, enhancing stability. However, despite these improvements, most of these models remain focused on natural images, with limited application to medical imaging.

\begin{table}[t]
\small
\centering
\caption{Comparison of the proposed CPDM with existing I2I translation methods. \textbf{S}, \textbf{R} and \textbf{F} stand for the \textbf{Sharpness}, \textbf{Realism}, and \textbf{Faithfulness} of the visual performance, estimated in three levels: High (H), Medium (M) and Low (L).} 
\label{tab:comparison_medical_i2i}
\resizebox{\linewidth}{!}
{%
\begin{tabular}{c|l|c|c|c|c|c}
    \toprule
    & \multirowcell{2}[-3pt][c]{\textbf{Method}}
    & \multirowcell{2}[-3pt][c]{\textbf{External} \\ \textbf{Knowledge}}
    & \multirowcell{2}[-3pt][c]{\textbf{Medical} \\ \textbf{I2I}}
    & \multicolumn{3}{c}{\textbf{Visual Per.}} 
    \\ 
    \cmidrule(lr){5-7} 
    &  &  &  &  \textbf{S} & \textbf{R} & \textbf{F} 
    \\ 
    \midrule
    \multirow{4}{*}{\rotatebox[origin=c]{90}{\textbf{GAN-based}}} & 
    Pix2Pix \cite{wang2018highresolutionimagesynthesissemantic} &  &  &  M & M & L
    \\ 
    & MedGAN~\cite{armanious2020medgan} &  & \cmark & M & M & M
    \\
    & UP-GAN \cite{zhu2017unpaired} &  & \cmark & M & M & M
    \\
    & FCN-cGAN\cite{bencohen2018crossmodalitysynthesisctpet} &  & \cmark & M & M & M \\
    \midrule
    \multirow{6}{*}{\rotatebox[origin=c]{90}{\textbf{Diffusion-based}}}
& Palette\cite{saharia2022palette} &  &  & H & H & L \\
    & DDIB\cite{su2022dual}  &  &  & H & H & L \\
    & SDEdit\cite{meng2021sdedit} & & & H & H & L \\
    &  LDM\cite{rombach2022high} & & & H & H & L \\
    & BBDM\cite{li2023bbdm} &  &  & H & H & L \\
   \cmidrule(lr){2-7} 
    & \textbf{CPDM} (Ours) & \cmark & \cmark & H & H & H \\
    \bottomrule
\end{tabular}
}
\end{table}
\noindent \textbf{Medical Image Translation Models.} 
In the medical domain, several GAN-based models have been developed for translating medical images, including PET-CT, MRI, and CT-to-CT tasks. 
MedGAN~\cite{armanious2020medgan} is one such framework that applies GANs for medical image translation, particularly in PET-CT translation, demonstrating the utility of GANs in clinical applications. 
Similarly, UP-GAN~\cite{zhu2017unpaired} adopts an uncertainty-guided progressive learning approach for translating between medical imaging modalities. 
Although these models have made significant strides in medical image translation, they are often limited by their reliance on paired data or struggle with generating high-fidelity images useful for diagnostic purposes.
In contrast, diffusion-based models have shown promise for overcoming the limitations of GANs in medical image translation. 
SynDiff~\cite{ozbey2023unsupervisedmedicalimagetranslation} utilizes a conditional diffusion process to generate high-quality medical images. 
However, its reliance on cycle-consistent architecture for training still poses inherent limitations, particularly in generating diagnostically relevant outputs.

To address current challenges, we propose CPDM, a novel diffusion-based method designed for CT2PET translation tasks. 
Unlike existing methods, CPDM leverages a Brownian Bridge process to directly synthesize PET images from CT images by incorporating domain-specific knowledge to guide the diffusion process, enhancing the quality and clinical relevance of the results.
We compare CPDM with existing I2I translation methods in Table~\ref{tab:comparison_medical_i2i}.

\section{Our Proposed PET/CT Dataset}
\label{sec:dataset}
\begin{table*}[t]
\renewcommand{\arraystretch}{1}
    \centering
    \small
    \caption{Detailed statistics of our large-scale CT-PET dataset.}
    \label{tab:statistics_ctpet}
    \resizebox{\linewidth}{!}
    {%
        \begin{tabular}{l|l||l|l||l|l}
            \toprule
            \multicolumn{2}{c||}{\textbf{General Information}} 
            & \multicolumn{2}{c||}{{\textbf{CT Information}}}
            & \multicolumn{2}{c}{{\textbf{PET Information}}} \\
            \midrule
            {Total CT-PET slices} & 2,028,628 
            & {Resolution (per image)} & 512 x 512 x 1 
            & {Resolution (per image)} & 256 x 256 x 1 
            \\
            {Total studies} & 3,454 
            & {Tube voltage (kVp)} & 120 / 140 
            & {Radioactive tracer} & $^{18}$F-FDG 
            \\
            {Total pairs per study} & 250-500
            & {Slice thickness (mm)} & 3.75 / 5 
            & {Slice thickness (mm)} & 3.27 
            \\
            {Body parts} & Whole body 
            & {Slope coefficient} & 1.0 
            & {Attenuation correction} & CT-based
            \\
            {PET/CT system} & GE Discovery 710 / STE
            & {Intercept coefficient} & -1024.0 
            & {Uptake time (minutes)} & 60
            \\
            \bottomrule
        \end{tabular}
`    }
\end{table*}

\textbf{Full Dataset.} 
We present a comprehensive large-scale PET/CT dataset collected from hospitals. 
The dataset consists of 2,028,628 paired CT-PET images from studies of 3,454 patients, designed to cover a wide range of anatomical regions. 
To ensure compliance with privacy and ethical guidelines, we remove all pathological labels rigorously. 
Each study includes approximately 250-500 paired CT and PET slices from the top of the head to the upper thigh region, above the knees. 
The images are stored in DICOM format, encapsulating pixel data and relevant metadata, including patient age (in years), sex, body weight, injected radiotracer activity, and other acquisition parameters.
Importantly, the PET images have undergone attenuation correction using the corresponding CT data. 
The images were acquired using the GE Discovery 710 PET/CT and GE Discovery STE PET/CT systems, with acquisition parameters defined by factors such as Kilovoltage peak (kVp), slope coefficient, and intercept coefficient. 
The dataset contains both diseased and non-diseased cases, providing a diverse representation of clinical scenarios. 
Table~\ref{tab:statistics_ctpet} provides the detailed statistics of this dataset.

\textbf{Experimental Dataset.}
Due to computational constraints, we limit our experiments (in Section \ref{sec:evaluation}) to a subset of the complete dataset. Specifically, we derive a sub-dataset of 30000 CT-PET images (15000 pairs), corresponding to 598 studies, by randomly selecting 25-30 paired slices per study. The experimental dataset is organized with sequential indexing at the patient level, i.e.,  images with consecutive indices correspond to contiguous anatomical sections from the same individual. The dataset is subsequently stratified into training, validation, and test sets using an 80:10:10 split. That is, we use 80\% of the studies allocated to the training set and the remaining 20\% of data for the validation and test sets. Consequently, the test set contains entirely unseen data during model training. All image pairs have been standardized to a resolution of $256 \times 256 \times 1$ and normalized to a range of $[-1; 1]$, based on the maximum pixel intensities of $2^{11}-1$ for CT images and $2^{15}-1$ for PET images. Detailed statistics and analysis of the experiment dataset are provided in Appendix 4.

\begin{figure*}[t]
    \centering
    \includegraphics[width=0.9\linewidth]{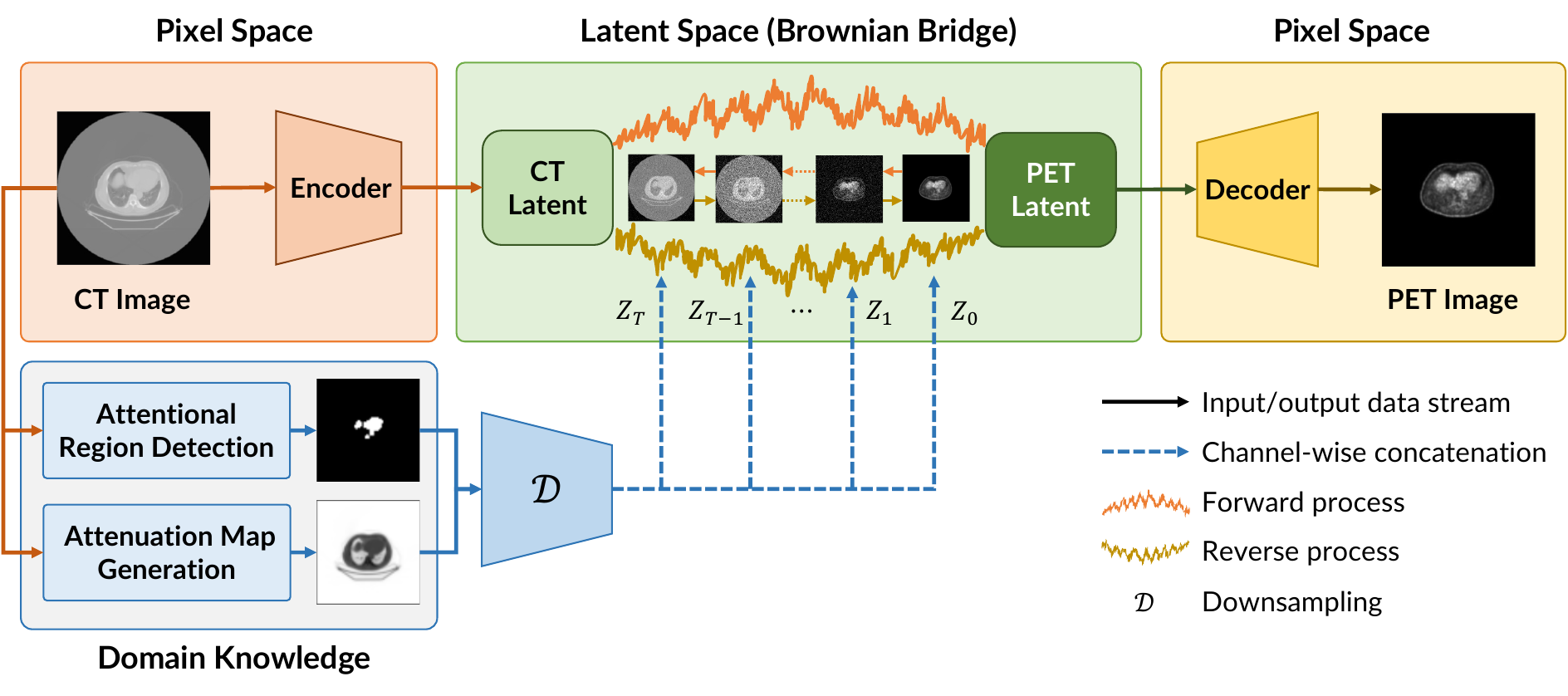}
     \vspace{-0.2cm}
    \caption{\textbf{Overview of CPDM.} 
    The medical knowledge Attention map and the Attenuation map are used to guide each stage of the Brownian Bridge diffusion process.}
    \label{fig:proposed_model}
\end{figure*}
\section{CPDM: Domain-Knowledge-Guided CT-to-PET Diffusion Model}
\label{sec:methodology}

\subsection{Motivation and Architecture Overview}
\label{subsec:motivation}
CT and PET images exhibit distinct characteristics compared to natural images. In these medical images, the majority of pixels are black, with only a small portion of pixels containing the critical non-black pixels that convey the most essential information. Specifically, in PET images, the intensely bright areas (High-SUV regions) are particularly significant, as they often signal regions with a high likelihood of abnormalities.
Translating CT images into PET images, thus, introduces unique challenges not encountered in natural image translation. 
First, because the bright regions are the most crucial, the translation process must prioritize these areas rather than distributing focus across the entire image. Second, the generated image must be sufficiently sharp to emphasize the critical bright regions effectively. Traditional generative models struggle to meet these specific requirements, as they tend to produce images with high variability, focusing more on fitting a general distribution than on achieving precise detail.

To address these challenges, we propose to use condition maps to guide the diffusion process, facilitating the generation of images that align with the inherent characteristics of PET images. For the first challenge, we introduce an attention map highlighting the regions likely to appear bright in the PET image. This attention map is derived from the CT image using a deep neural network trained through supervised learning. To tackle the second challenge,  we based on the fact that different body parts absorb radiation at varying levels. We, thus, employ an attenuation map, a matrix that represents the absorption levels of radiation (light intensity) across different body parts
The attenuation map can be accurately calculated from the input CT image. Indeed, the attenuation map is traditionally used to correct PET images, making it an ideal tool for enhancing image clarity and focus.
Additionally, inspired by~\cite{li2023bbdm}, we integrate the Brownian Bridge process into our diffusion model for CT2PET translation. 
In contrast to existing diffusion methods, the proposed approach directly establishes a mapping between the input CT and output PET domains. This direct mapping eliminates the stochasticity typically associated with generative models. 
%

Figure~\ref{fig:proposed_model} illustrates the architecture of the proposed method, named CPDM. Starting with a CT image from the CT domain, we first extract its latent feature $L_\text{CT}$. Following this, the Brownian Bridge process maps 
$L_\text{CT}$ to its corresponding latent representation $L_{\text{CT} \to \text{PET}}$ in the PET domain. During this  process, domain knowledge in the form of Attention map ($\boldsymbol{M_\sigma}$) and Attenuation map ($\boldsymbol{M_\mu}$) are employed. Subsequently, the synthesized PET image is generated by the decoder. We leverage a pre-trained VQGAN~\cite{esser2021taming} for the Encoder/Decoder on our PET/CT dataset to enhance learning efficiency and model generalization.

\subsection{Domain Knowledge-Guided Maps}
\label{subsec:maps}
\textbf{Attention Map ($\boldsymbol{M_\sigma}$).} 
The Attention map identifies clinical regions of interest in CT images that show increased $^{18}$F-Fluorodeoxyglucose ($^{18}$F-FDG) and exhibit higher SUVs uptake in PET images. 
To generate this map, we employ the U-Net~\cite{ronneberger2015u} segmentation model, which is trained on CT images to distinguish these high-activity tissues, using Dice Loss~\cite{milletari2016v} function
 to measure accuracy. 
PET images are employed to define the ground truth for the U-Net model by creating segmented masks that highlight areas of interest based on their brightness. In detail, we perform a qualitative assessment of PET images across different anatomical regions and utilize a straightforward thresholding approach to sample segmented masks, isolating pixels with intensity values that exceed a predefined threshold. The loss function is defined as follows:

\scalebox{0.9}{
\begin{minipage}{0.52\textwidth}
\begin{equation*}
\mathcal{L}_{\text{$\boldsymbol{M_\sigma}$}} = 1 - \frac{2 |M_{pre} \cap  M_{gt}|}{|M_{pre}| + |M_{gt}|},
\end{equation*}
\end{minipage}
}

where $M_{pre}$ denotes the predicted mask from segmentation model and $M_{gt}$ represents the ground truth mask. The resulting Attention map $\boldsymbol{M_\sigma}$ then guides the Brownian Bridge diffusion process to focus on the highlighted regions when generating the PET images. 

\noindent \textbf{Attenuation Map ($\boldsymbol{M_\mu}$).}
In practical settings, PET/CT scanners typically utilize CT images to correct attenuation, thereby improving the quality of PET emission data. 
This crucial process involves converting CT data, acquired across a broad spectrum of photon energies (approximately 30 to 140 kVp), into attenuation coefficients for PET photon energies at 511 keV.
Inspired by~\cite{abella2012accuracy}, we employ the tri-linear scaling methodology to calculate the Attenuation map.
The Linear Attenuation Coefficient (LAC) at 511 keV, denoted as $\mu$, is calculated using the equation $\mu = \alpha \times HU + \beta$, where $\alpha$ and $\beta$ are constants derived from scanner calibration, which are listed in~\cite{abella2012accuracy}, and $HU$ represents the Hounsfield Units calculated from CT data (i.e. slope coefficient, intercept coefficient, and X-ray tube potential).
The generated Attenuation map $\boldsymbol{M_\mu} = e^{-\mu}$ is subsequently incorporated into the Brownian Bridge diffusion process, providing a guidance informed by medical expertise for model refinement. 

\subsection{Conditional Brownian Bridge Diffusion Process}
\label{subsec:diffusion process}
Drawing inspiration from \cite{li2023bbdm}, we adopt the Brownian Bridge process to characterize the probability distribution throughout the diffusion. This process aims to facilitate a direct transformation from the source to the target domain. 
To enhance the training process's efficiency, we use an encoder and decoder to transform PET and CT images into latent space vectors before the diffusion process. Denoting the original PET and CT images as $\boldsymbol{X}$ and $\boldsymbol{Y}$, and the encoder and decoder as $\mathcal{E}$ and $\mathcal{D}$, the corresponding latent vectors, $\boldsymbol{x}$ and $\boldsymbol{y}$, are obtained as $\boldsymbol{x} := \mathcal{E}(\boldsymbol{X})$ and $\boldsymbol{y} := \mathcal{E}(\boldsymbol{Y})$. During the training phase, CPDM learns a mapping from $\boldsymbol{y}$ to $\boldsymbol{x}$ in the latent space through a Brownian Bridge process. At the end of the inference phase, the translated PET image $\tilde{\boldsymbol{X}}$ is generated by the decoder of the pre-trained VQGAN as $\tilde{\boldsymbol{X}} := \mathcal{D}(\tilde{\boldsymbol{x}})$, where $\tilde{\boldsymbol{x}}$ is the final latent state of the reverse diffusion process.

\noindent \textbf{Forward Process.}
We utilize the formulation proposed in \cite{li2023bbdm}.
Starting from an initial state $\boldsymbol{x}_0$ (i.e., the latent vector of a PET image) and aiming for a destination state $\boldsymbol{y}$ (i.e., the latent vector of the corresponding CT image), the intermediate state $\boldsymbol{x}_t$ at timestep $t$ are determined in discrete form as $\boldsymbol{x}_t=\left(1-m_t\right) \boldsymbol{x}_0+m_t \boldsymbol{y}+\sqrt{\delta_t} \epsilon_t$, where $m_t = t / T$, $T$ represents the total number of steps in the diffusion process, $\delta_t$ indicates the variance of the Brownian Bridge, and $\epsilon$ depicts a Gaussian noise, i.e., $\epsilon \sim \mathcal{N}(\mathbf{0}, \mathbf{I})$.
The forward process is defined as: \\
\scalebox{0.9}{
\begin{minipage}{0.52\textwidth}
\begin{equation}
    q_{BB} \left( \boldsymbol{x}_t | \boldsymbol{x}_0, \boldsymbol{y}) = \mathcal{N} (\boldsymbol{x}_t; (1 - m_t) \boldsymbol{x}_0 + m_t \boldsymbol{y}, \delta_t \mathbf{I} \right).
\end{equation}
\end{minipage}
}

Throughout the training phase, we employ the following formula to establish the transition probability between two consecutive steps: \\
\scalebox{0.9}{
\begin{minipage}{0.52\textwidth}
\begin{align}
    q_{BB} (\boldsymbol{x}_t | \boldsymbol{x}_{t-1}, \boldsymbol{y}) &= \mathcal{N} \left( \boldsymbol{x}_t;  \frac{1 - m_t}{ 1 - m_{t-1} } \boldsymbol{x}_{t-1} \right. \nonumber \\ 
    &\left. + \left(m_t - \frac{1 - m_t}{ 1 - m_{t-1}} m_{t-1}\right) \boldsymbol{y}, \delta_{t|t-1} \mathbf{I} \right), \nonumber \\
    \text{with} \quad \delta_{t \mid t-1} &= \delta_t - \delta_{t-1} \frac{\left(1 - m_t\right)^2}{\left(1 - m_{t-1}\right)^2}.
\label{eq:forward_process}
\end{align}
\end{minipage}
}
\begin{algorithm}[t]
\caption{CPDM Training Process}\label{alg:training_proposed_model}
\begin{algorithmic}[1]
\small
\REPEAT
    \STATE Paired data: PET image $\boldsymbol{x_0} \sim q(\boldsymbol{x_0})$, CT image $\boldsymbol{y} \sim q(\boldsymbol{y})$;
    Attention map $\boldsymbol{M_\sigma}$; Attenuation map $\boldsymbol{M_\mu}$ 
    \STATE Timestep \scalebox{0.9}{$t \sim Uniform(1, \dots, T)$}
    \STATE Gaussian noise \scalebox{0.9}{$\epsilon \sim \mathcal{N}(\mathbf{0}, \mathbf{I})$}
    \STATE Forward diffusion \scalebox{0.9}{$\boldsymbol{x_t} =\left(1-m_t\right) \boldsymbol{x_0} +  m_t  \boldsymbol{y}+\sqrt{\delta_t} \epsilon$}
    \STATE Take gradient descent: \\
     \scalebox{0.9}{$\nabla_\theta\left\|m_t\left(\boldsymbol{y} - \boldsymbol{x_0} \right)+\sqrt{\delta_t} \epsilon - \epsilon_\theta \left(\text{concat}(\boldsymbol{x_t}, \boldsymbol{M_\sigma}, \boldsymbol{M_\mu}), t\right)\right\|_1$}
     \normalsize
\UNTIL{converged}
\normalsize
\end{algorithmic}
\end{algorithm}

\noindent \textbf{Reverse Process.} 
The reverse initiates by conditionally assigning $\boldsymbol{x}_T = \boldsymbol{y}$. Furthermore, to incorporate the domain knowledge, we employ the channel-wise concatenation to integrate the Attention map $\boldsymbol{M}_\sigma$ and the Attenuation map $\boldsymbol{M}_\mu$ within the hidden state $\boldsymbol{x}_t$ at every timestep $t$ as: \\
\scalebox{0.86}{
\begin{minipage}{0.55\textwidth}
\begin{align}
    p_\theta(\boldsymbol{x}_{t-1} | \boldsymbol{x}_t, \boldsymbol{M}_\sigma, \boldsymbol{M}_\mu, \boldsymbol{y}) =&~ \mathcal{N} (\boldsymbol{x}_{t-1}; \mu_\theta (\boldsymbol{x}_t, \boldsymbol{M}_\sigma, \boldsymbol{M}_\mu, \boldsymbol{y}, t ), \tilde{\delta}_t \mathbf{I}), \nonumber \\
   \mu_\theta (\boldsymbol{x}_t, \boldsymbol{M}_\sigma, \boldsymbol{M}_\mu, \boldsymbol{y}, t) =&~ c_{x t} \boldsymbol{x}_t + c_{y t} \boldsymbol{y} \nonumber \\ 
   + & c_{\epsilon t} \boldsymbol{\epsilon}_\theta (\text{concat}(\boldsymbol{x}_t, \boldsymbol{M}_\sigma, \boldsymbol{M}_\mu), t), 
\label{eq:reverse_process}
\end{align}
\end{minipage}
}

where $\mu_\theta (\boldsymbol{x}_t, \boldsymbol{M}_\sigma, \boldsymbol{M}_\mu, \boldsymbol{y}, t)$ represents the predicted mean and $\tilde{\delta}_t$ represents the variance of the distribution at time step $t$, respectively; $c_{x t}$, $c_{y t}$ and $c_{\epsilon t}$ are non-trainable factors derived from $m_t$, $m_{t-1}$, $\delta_t$ and $\delta_{t-1}$:\\
\scalebox{0.9}{
\begin{minipage}{0.52\textwidth}
\begin{equation}
    \begin{aligned}
        c_{x t} &= \frac{\delta_{t-1}}{\delta_t} \frac{1-m_t}{1-m_{t-1}}+\frac{\delta_{t \mid t-1}}{\delta_t}\left(1-m_{t-1}\right), \\
        c_{y t} &= m_{t-1}-m_t \frac{1-m_t}{1-m_{t-1}} \frac{\delta_{t-1}}{\delta_t}, 
        \quad c_{\epsilon t} = \left(1-m_{t-1}\right) \frac{\delta_{t \mid t-1}}{\delta_t}. \nonumber
    \end{aligned}
\end{equation}
\end{minipage}
}

\noindent \textbf{Training Objective.} 
The training process entails optimizing the disparity between the predicted distribution of the latent variables and the observed distribution in the forward diffusion process (see Algorithm~\ref{alg:training_proposed_model}). This is achieved by learning the mean value $\mu_\theta \left(\boldsymbol{x}_t, \boldsymbol{M}_\sigma, \boldsymbol{M}_\mu, \boldsymbol{y}, t\right)$ through a neural network parameterized by $\theta$, employing maximum likelihood estimation. In practice, this is accomplished by minimizing the following Evidence Lower Bound (ELBO):
%
\scalebox{0.86}{
\begin{minipage}{0.55\textwidth}
\centering
\begin{equation*}
    \begin{aligned}
        &\text{ELBO} = -\mathbb{E}_q\left(D_{K L}\left(q_{B B}\left(\boldsymbol{x}_T \mid \boldsymbol{x}_0, \boldsymbol{y}\right) \|  
        p\left(\boldsymbol{x}_T \mid \boldsymbol{M}_\sigma, \boldsymbol{M}_\mu, \boldsymbol{y}\right)\right)\right. \\
        & +\sum_{t=2}^T D_{K L}\left(q_{B B}\left(\boldsymbol{x}_{t-1} \mid \boldsymbol{x}_t, \boldsymbol{x}_0, \boldsymbol{y}\right) \| p_\theta\left(\boldsymbol{x}_{t-1} \mid \boldsymbol{x}_t, \boldsymbol{M}_\sigma, \boldsymbol{M}_\mu, \boldsymbol{y}\right)\right)  
         \\
        & \left.-\log p_\theta\left(\boldsymbol{x}_0 \mid \boldsymbol{x}_1, \boldsymbol{M}_\sigma, \boldsymbol{M}_\mu, \boldsymbol{y}\right)\right).
    \end{aligned}
\end{equation*}
\end{minipage}
}

By substituting the values of $q_{B B}(.)$ and $p_{\theta}(.)$ from Equations~\ref{eq:forward_process} and ~\ref{eq:reverse_process}, respectively, we derive the final training ELBO objective as follows:\\
\scalebox{0.82}{
\begin{minipage}{0.5\textwidth}
\begin{equation}
\nonumber
    \mathbb{E}_{\boldsymbol{x}_0, \boldsymbol{y}, \boldsymbol{\epsilon}}\left[c_{\epsilon t}\left\|m_t\left(\boldsymbol{y}-\boldsymbol{x}_0\right)+\sqrt{\delta_t} \epsilon - \epsilon_\theta\left(\text{concat}(\boldsymbol{x}_t, \boldsymbol{M}_\sigma, \boldsymbol{M}_\mu), t\right)\right\|_1\right].
\end{equation}
\end{minipage}
}

\noindent \textbf{Sampling Process.} We adopt the basic idea of DDIM (Denoising Diffusion Implicit Models)~\cite{song2020denoising}, where acceleration is achieved through a non-Markovian process while maintaining identical marginal distributions as Markovian inference processes (see Algorithm~\ref{alg:sampling_proposed_model}).

\begin{algorithm}[t]
\caption{CPDM Sampling Process}\label{alg:sampling_proposed_model}
\begin{algorithmic}[1]
\small
\STATE Sample conditional input: CT image $\boldsymbol{x_T} = \boldsymbol{y} \sim q(\boldsymbol{y})$; \\ Attention map $\boldsymbol{M_\sigma}$; Attenuation map $\boldsymbol{M_\mu}$ 
\FOR{$t=T, \ldots, 1$}
    \STATE \scalebox{0.86}{$\boldsymbol{z} \sim \mathcal{N}(\mathbf{0}, \mathbf{I})$ if $t>1$, else $\boldsymbol{z} = \mathbf{0}$}
    \STATE \scalebox{0.86}{$\boldsymbol{x_{t-1}} =  c_{xt} \boldsymbol{x_t} + c_{yt}\boldsymbol{y} -c_{\epsilon t} \epsilon_\theta\left(\text{concat}(\boldsymbol{x_t}, \boldsymbol{M_\sigma}, \boldsymbol{M_\mu}), t\right)+\sqrt{\tilde{\delta}_t} \boldsymbol{z} $}
\ENDFOR
\RETURN $\boldsymbol{x_0}$
\normalsize
\end{algorithmic}
\end{algorithm}

\section{Performance Evaluation}
\label{sec:evaluation}
In this section, we evaluate the performance of CPDM using our PET/CT dataset. We compare CPDM  against SOTA image translation methods, including GAN-based models, i.e., Pix2Pix~\cite{isola2017image}, MedGAN~\cite{armanious2020medgan}, and UP-GAN~\cite{upadhyay2021uncertainty} and diffusion-based models, i.e., DDIB~\cite{su2022dual}, Palette~\cite{saharia2022palette}, SDEdit~\cite{meng2021sdedit}, LDM~\cite{rombach2022high}, and BBDM~\cite{li2023bbdm} (Section~\ref{sec:result}). We also provide the ablation studies in Section~\ref{sec:ablation}.


\subsection{Experimental Settings}
\label{sec:setting}
\textbf{Evaluation Metrics.}
Following~\cite{armanious2020medgan, upadhyay2021uncertainty}, we utilize Learned Perceptual Image Patch Similarity (LPIPS), Mean Absolute Error (MAE), Structural Similarity Index Measure (SSIM), and Peak Signal-to-Noise Ratio (PSNR) metrics to evaluate the quality of generated PET images (detailed in Appendix 2.3). These metrics ensure that the generated PET images are accurate and of high visual quality. LPIPS evaluates perceptual similarity, which is crucial for capturing medical details. MAE measures the average pixel-wise error, indicating accuracy. SSIM reflects visual quality by assessing changes in the image structure, whereas PSNR quantifies pixel-level fidelity. 

\textbf{Models \& Hyperparameters.}
We implement the CPDM as detailed in Section~\ref{sec:methodology}. The diffusion timesteps of the Brownian Bridge are set at $1000$ for training and reduced to $200$ for inference, aiming to balance quality and speed. The Unet~\cite{ronneberger2015u}, utilizing a pre-trained ResNet50~\cite{he2016deep} as its backbone, is employed for extracting the attention map. The hyperparameters of the baselines remain consistent with those specified in the original papers (detailed in Appendix 2.2). We adjust the input and output image channels to $1$ and a batch size to $16$ to optimize hardware processing capabilities. We utilize Adam~\cite{kingma2014adam} optimizer with an initial learning rate of \(10^{-4}\), with the step decay for learning rate scheduling and a batch size of 16 across 200 epochs. 

\begin{table}[t]
    \centering
    \small
    \caption{\textbf{Comparison of CPDM against SOTA methods.} The best and second-based results are highlighted by \textbf{\textcolor{red}{red}} and \textbf{\textcolor{blue}{blue}}, respectively. \textbf{Diff.} shows the relative performance gaps of CPDM compared to the nearest methods. $\downarrow$/$\uparrow$ means lower/higher is better.} 
    \label{tab:comparison}
    \resizebox{\linewidth}{!}
    {%
        \begin{tabular}{l|rrrc}
            \toprule
            \textbf{Method} & \textbf{LPIPS $\downarrow$} & \textbf{MAE $\downarrow$} & \textbf{SSIM $\uparrow$} & \textbf{PSNR $\uparrow$} \\
            \midrule
            Pix2Pix~\cite{isola2017image} & \second{0.0491} & 336.47 & 0.9298 & 27.52 \\
            MedGAN~\cite{armanious2020medgan} & 0.1322 & 336.97 & 0.9329 & \second{29.27} \\
            UP-GAN~\cite{upadhyay2021uncertainty} & 0.1082 & 373.49 & \second{0.9337} & 28.38 \\ 
            \midrule
            DDIB~\cite{su2022dual} & 0.5556 & 4898.58 & 0.0343 & 16.10 \\
            Palette~\cite{saharia2022palette} & 0.3827 & 4032.23 & 0.3054 & 20.92 \\
            SDEdit~\cite{meng2021sdedit} & 0.1577 & 813.65 & 0.8187 & 22.85 \\
            LDM~\cite{rombach2022high} & 0.0492 & \second{318.71} & 0.9317 & 28.30 \\
            BBDM~\cite{li2023bbdm} & 0.0548 & 353.66 & 0.9232 & 27.56 \\ \cmidrule(lr){1-5}
            \textbf{CPDM} (w/o $M_\sigma$) & 0.0471 & 305.46 & 0.9353 & 28.76 \\
            \textbf{CPDM} (w/o $M_\mu$) & 0.0523 & 324.99 & 0.9277 & 28.47 \\
            \textbf{CPDM} (Ours) & \best{0.0466} & \best{284.61} & \best{0.9396} & \best{29.68} \\
            \midrule
            \textbf{Diff.} &  \fcolorbox{green!25}{green!25}{-5.1\%} &  \fcolorbox{green!25}{green!25}{-10.7\%} &  \fcolorbox{green!25}{green!25}{+0.6\%} & \fcolorbox{green!25}{green!25}{+1.4\%} \\
            \bottomrule
        \end{tabular}
    }             
    \vspace{-0.2cm}
\end{table}

\begin{figure*}[t]
    \centering
    \begin{minipage}[t]{0.67\textwidth} 
        \centering
        \includegraphics[width=\textwidth]{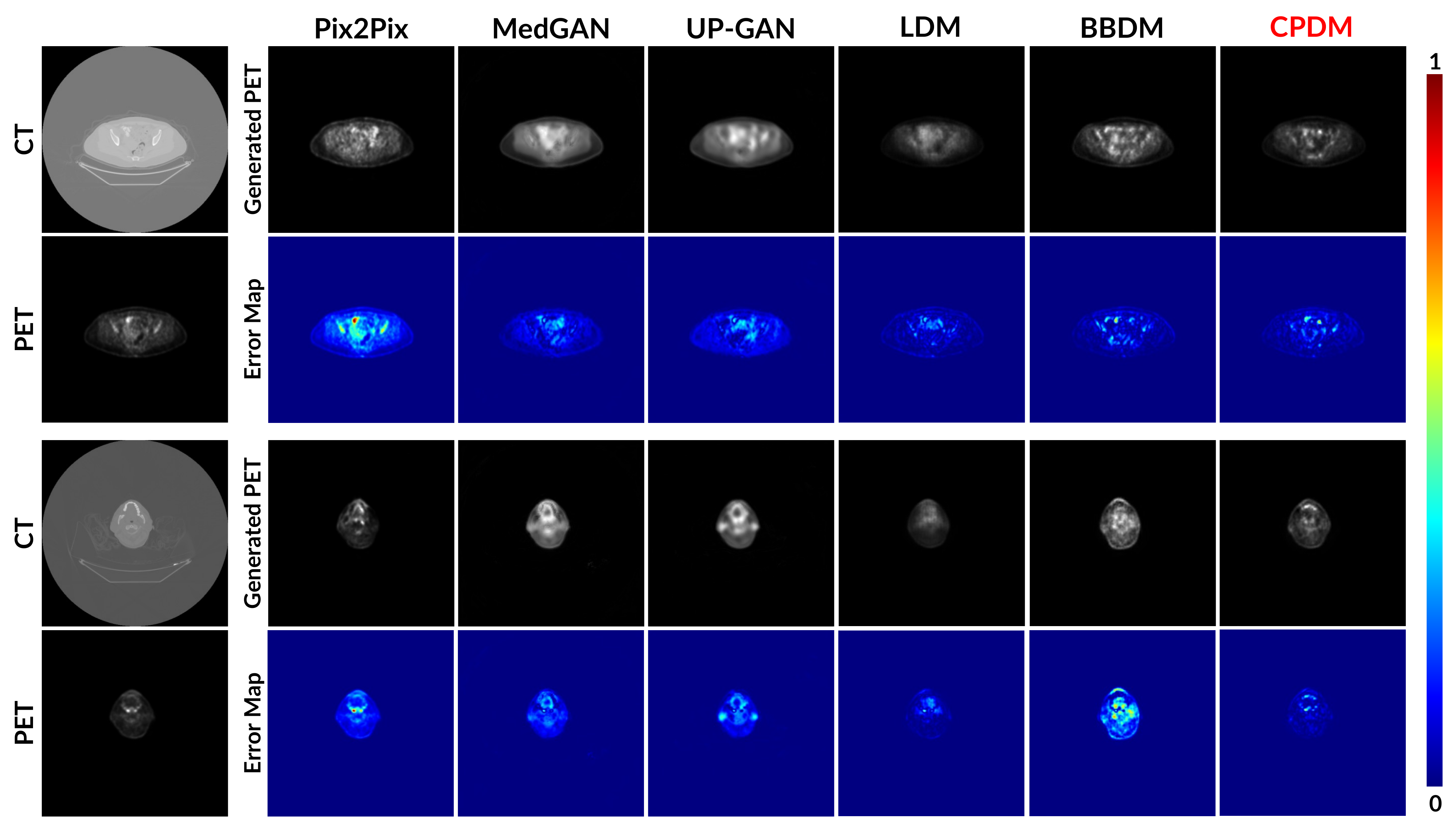}
        \caption{\textbf{Visualization of PET images generated by the best-performing methods.} Errors induced by PET produced by CPDM are significantly more minor than other methods.}  
        \label{fig:result_visualization}
    \end{minipage}%
    \hfill 
    \begin{minipage}[t]{0.31\textwidth}
        \centering
        \includegraphics[width=\textwidth]{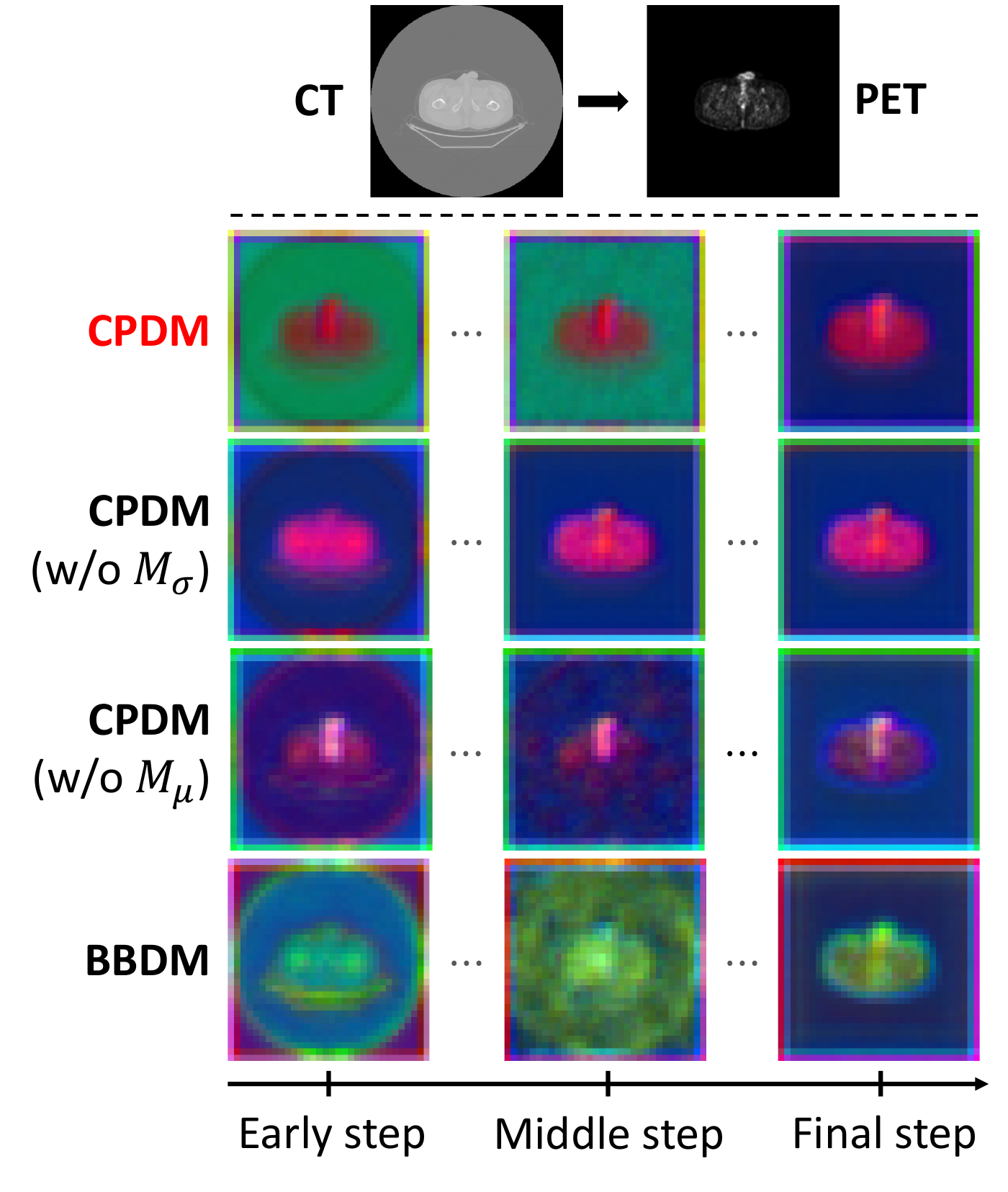}
        \caption{
        {
        Visualization of spatial features extracted from the denoising network $\epsilon_\theta$.} 
        }
        \label{fig:pca_visualization}
    \end{minipage}
\end{figure*}

\subsection{Comparison with Existing Methods}
\label{sec:result}
Table \ref{tab:comparison} presents a thorough comparison of PET image generation models, where CPDM stands out by achieving the best results across all metrics: lowest LPIPS (0.0466), lowest MAE (284.61), highest SSIM (0.9396), and highest PSNR (29.68). It demonstrates that CPDM provides superior perceptual quality and structural fidelity. Other methods perform well on specific metrics but often underperform on others. 
For instance, MedGAN achieves the second-best PSNR (29.27) but ranks fourth-worst in LPIPS (0.1322), highlighting its trade-off between pixel-level noise reduction and perceptual quality. Similarly, UP-GAN performs well in SSIM but falls behind in LPIPS. Overall, CPDM offers a more balanced and robust performance across all metrics. The final row (Diff.) shows CPDM's improvement over the second-best models, with a 5.1\% reduction in LPIPS, a 10.7\% decrease in MAE, a 0.6\% increase in SSIM, and a 1.4\% rise in PSNR.

To further investigate the differences between CPDM and other methods, we visualize several examples in \figref{fig:result_visualization}.
PET images generated by CPDM exhibit sharper details than those produced by the other models. As error maps show, most of CPDM's pixels appear dark blue, indicating minimal reconstruction error. In contrast, GAN-based models, such as MedGAN and UP-GAN, demonstrate relatively good faithfulness to the input data but often suffer from insufficient sharpness and lack of realism in the generated images. Conversely, diffusion-based models like LDM, which rely on conditional synthetic generation, produce images with high sharpness and detailed structures; however, they exhibit lower faithfulness, struggling to preserve anatomical integrity accurately. BBDM achieves a more balanced performance, offering a reasonable trade-off between sharpness, faithfulness, and realism. However, it lacks the domain-specific knowledge necessary to match the anatomical accuracy and structural faithfulness that CPDM consistently delivers. It highlights CPDM's ability to maintain a superior balance across crucial image quality attributes, outperforming other methods.

\subsection{Ablation Studies}
\label{sec:ablation}
\subsubsection{Impacts of Domain Knowledge Guidance}
We implement two variants of CPDM, namely CPDM (w/o $\boldsymbol{M_\sigma}$) and CPDM (w/o $\boldsymbol{M_\mu}$), to examine the impacts of the two proposed conditional maps. 



\noindent \textbf{Quantitative Results.}
The results in Table~\ref{tab:comparison} show that both variants of CPDM perform worse than the original CPDM across all metrics, underscoring the effectiveness of utilizing both conditional maps.
Specifically, omitting the Attention map~$\boldsymbol{M_\sigma}$ leads to an increase of approximately $1.1\%$ and $7.3\%$ in LPIPS and MAE, along with decreases of $0.46\%$ and $3.0\%$ in SSIM and PSNR, respectively. In contrast, removing the Attenuation map $\boldsymbol{M_\mu}$ results in significant increases of $12.2\%$ and $14.2\%$ in LPIPS and MAE, accompanied by decreases of $1.3\%$ and $4.1\%$ in SSIM and PSNR, respectively. The results demonstrate that the Attenuation map~$\boldsymbol{M_\mu}$ is more critical than the Attention map~$\boldsymbol{M_\sigma}$, as its absence causes more substantial degradation across all metrics.


\noindent \textbf{Qualitative Results.}
To evaluate the effects of the Attention map $\boldsymbol{M_\sigma}$ and the Attenuation map $\boldsymbol{M_\mu}$ on features generated by the diffusion process, we use Principal Component Analysis (PCA)~\cite{dunteman1989principal} to visualize the intermediate spatial features throughout the denoising network $\epsilon_\theta$.
\figref{fig:pca_visualization} shows that the BBDM model spreads features into the surrounding background rather than focusing on patient anatomy.
In contrast, the proposed CPDM effectively integrates both two condition maps, achieving focused attention on regions of interest and meaningful anatomical structures.
The CPDM without $\boldsymbol{M_\sigma}$ (second row) shows an enhanced focus on anatomical features during the diffusion process, particularly at intermediate timesteps ($t \approx T / 2$), where it maintains consistency even in high noise conditions.
Meanwhile, CPDM without $\boldsymbol{M_\mu}$ (third row) demonstrates that the Attention map helps $\epsilon_\theta$ focus on significant features related to regions of interest, maintaining this focus even at intermediate timesteps with minimal information.

\subsubsection{Robustness of the Condition Maps}
\begin{table}[t]
\small
\renewcommand{\arraystretch}{0.6}
    \centering
    \caption{Impact of the our conditional maps on existing methods.} 
    \label{tab:plugin_comparison}
    \resizebox{\linewidth}{!}
    {%
        \begin{tabular}{l|l||rrrc}
            \toprule
            \multicolumn{2}{c||}{\textbf{Method} }&\textbf{LPIPS $\downarrow$} & \textbf{MAE $\downarrow$} & \textbf{SSIM $\uparrow$} & \textbf{PSNR $\uparrow$} \\
            \midrule 
            Pix2Pix & Original & 0.0491 & 336.47 & 0.9298 & 27.52 \\ \cmidrule(lr){2-6}
             & w/ ($M_\sigma, M_\mu$) & 0.0470 & 305.33 & 0.9345 & 28.49 \\ \cmidrule(lr){2-6} 
            & \textbf{Diff.}  & \fcolorbox{green!25}{green!25}{-4.3\%} & \fcolorbox{green!25}{green!25}{-9.3\%} & \fcolorbox{green!25}{green!25}{+0.50\%} & \fcolorbox{green!25}{green!25}{+3.4\%} \\ \midrule
            MedGAN & Original & 0.1322 & 336.97 & 0.9329 & 29.27 \\ \cmidrule(lr){2-6}
             & w/ ($M_\sigma, M_\mu$) & 0.1158 & \second{294.05} & 0.9325 & \second{29.73} \\ \cmidrule(lr){2-6} 
            & \textbf{Diff.} & \fcolorbox{green!25}{green!25}{-12.4\%} & \fcolorbox{green!25}{green!25}{-12.7\%} & \fcolorbox{red!25}{red!25}{-0.04\%} & \fcolorbox{green!25}{green!25}{+1.5\%} \\ \midrule
            UP-GAN & Original & 0.1082 & 373.49 & 0.9337 & 28.38 \\ \cmidrule(lr){2-6} 
             & w/ ($M_\sigma, M_\mu$) & 0.1001 & 301.67 & 0.9296 & \best{29.99} \\ \cmidrule(lr){2-6} 
            & \textbf{Diff.} & \fcolorbox{green!25}{green!25}{-7.5\%} & \fcolorbox{green!25}{green!25}{-19.2\%} & \fcolorbox{red!25}{red!25}{-0.44\%} & \fcolorbox{green!25}{green!25}{+5.7\%} \\ \midrule
            LDM & Original & 0.0492 & 318.71 & 0.9317 & 28.30 \\ \cmidrule(lr){2-6}
            & w/ ($M_\sigma, M_\mu$) & \second{0.0468} & 302.70  & \second{0.9348} & 28.92 \\ \cmidrule(lr){2-6} 
            & \textbf{Diff.} & \fcolorbox{green!25}{green!25}{-4.9\%} & \fcolorbox{green!25}{green!25}{-8.2\%} & \fcolorbox{green!25}{green!25}{+0.33\%} & \fcolorbox{green!25}{green!25}{+2.2\%} \\ 
            \midrule
            \multicolumn{2}{l||}{\textbf{CPDM} (Ours)} & \best{0.0466} & \best{284.61} & \best{0.9396} & 29.68 \\
            \bottomrule
        \end{tabular}
    }              
\end{table}
In this section, we evaluate the robustness of the proposed condition maps, $M_\sigma$, and $M_\mu$, by integrating them into various I2I models (i.e., Pix2Pix, MedGAN, UP-GAN, and LDM) and analyzing their impact on performance enhancement.
Specifically, we developed a conditional variant for each model that incorporates the proposed maps to guide the generative process, embedding domain-specific knowledge to refine model outputs.
Table~\ref{tab:plugin_comparison} presents quantitative results demonstrating that integrating our conditional maps improves the performance of the original methods.
The inclusion of \(M_\sigma\) and \(M_\mu\) leads to significant enhancements in LPIPS and MAE, with UP-GAN achieving gains of 7.5\% and 19.2\%, respectively. 
The slight reduction in MAE can be attributed to the conditional maps, which provide additional guidance during generation and effectively reduce pixel-wise discrepancies by embedding domain-specific information.
Despite these improvements, UP-GAN and MedGAN still exhibit considerably higher LPIPS than CPDM, indicating lower perceptual quality.
The conditional maps encourage these models to focus on faithfulness, resulting in smoother images with fewer pixel-level discrepancies but diminished sharpness and detail.
This trade-off likely explains the elevated PSNR and slight decline in SSIM observed in UP-GAN and MedGAN, as the generated images prioritize smoothness at the expense of visual fidelity.
In contrast, LDM, leveraging the gradual refinement process typical of diffusion models, shows more balanced improvements across all metrics by focusing on local and global consistency.
Nonetheless, CPDM surpasses LDM by achieving a superior balance between sharpness, realism, and faithfulness, preserving finer details while delivering better overall performance.


\section{Conclusion}
\label{sec:conclusion}
This study proposed CPDM, a novel method for synthesizing PET images from CT images using a diffusion process. 
CPDM utilized a Brownian Bridge diffusion process to translate data directly from the CT domain to the PET domain, mitigating the inherent stochastic characteristics often present in generative models. 
Furthermore, CPDM integrated Attention and Attenuation maps, resulting in PET images with significantly improved quality and reduced error compared to existing methods.
The quantitative and qualitative results demonstrated the superiority of CPDM over existing GAN-based and diffusion-based methods across various evaluation criteria. 

For future work, we plan to train and evaluate CPDM on the full dataset and collaborate with medical professionals to assess the clinical performance and diagnostic utility of the PET images generated by our approach.

\section*{Acknowledgment}
This work was funded by Vingroup Joint Stock Company (Vingroup JSC), Vingroup, and supported by Vingroup Innovation Foundation (VINIF) under project code VINIF.2021.DA00128.

{\small
\bibliographystyle{ieee_fullname}
\bibliography{egbib}

\begin{thebibliography}{10}\itemsep=-1pt

\bibitem{abella2012accuracy}
Monica Abella, Adam~M Alessio, David~A Mankoff, Lawrence~R MacDonald, Juan~Jose Vaquero, Manuel Desco, and Paul~E Kinahan.
\newblock Accuracy of {CT}-based attenuation correction in {PET/CT} bone imaging.
\newblock {\em Physics in Medicine \& Biology}, 57(9):2477, 2012.

\bibitem{alameen2021radiobiological}
Suhaib Alameen, Nissren Tamam, Sami Awadain, Abdelmoneim Sulieman, Latifa Alkhaldi, and Amira~Ben Hmed.
\newblock Radiobiological risks in terms of effective dose and organ dose from 18f-fdg whole-body pet/ct procedures.
\newblock {\em Saudi Journal of Biological Sciences}, 28:5947--5951, 2021.

\bibitem{armanious2020medgan}
Karim Armanious, Chenming Jiang, Marc Fischer, Thomas K{\"u}stner, Tobias Hepp, Konstantin Nikolaou, Sergios Gatidis, and Bin Yang.
\newblock Medgan: Medical image translation using {GANs}.
\newblock {\em Computerized Medical Imaging and Graphics}, 79:101684, 2020.

\bibitem{Ben_Cohen_2017}
Avi Ben-Cohen, Eyal Klang, Stephen~P. Raskin, Michal~Marianne Amitai, and Hayit Greenspan.
\newblock {\em Virtual PET Images from CT Data Using Deep Convolutional Networks: Initial Results}, page 49–57.
\newblock Springer International Publishing, 2017.

\bibitem{bencohen2018crossmodalitysynthesisctpet}
Avi Ben-Cohen, Eyal Klang, Stephen~P. Raskin, Shelly Soffer, Simona Ben-Haim, Eli Konen, Michal~Marianne Amitai, and Hayit Greenspan.
\newblock Cross-modality synthesis from ct to pet using fcn and gan networks for improved automated lesion detection, 2018.

\bibitem{cherry2018total}
Simon~R Cherry, Terry Jones, Joel~S Karp, Jinyi Qi, William~W Moses, and Ramsey~D Badawi.
\newblock Total-body {PET}: maximizing sensitivity to create new opportunities for clinical research and patient care.
\newblock {\em Journal of Nuclear Medicine}, 59(1):3--12, 2018.

\bibitem{croitoru2023diffusion}
Florinel-Alin Croitoru, Vlad Hondru, Radu~Tudor Ionescu, and Mubarak Shah.
\newblock Diffusion models in vision: A survey.
\newblock {\em IEEE Transactions on Pattern Analysis and Machine Intelligence}, 2023.

\bibitem{dhariwal2021diffusion}
Prafulla Dhariwal and Alexander Nichol.
\newblock Diffusion models beat {GANs} on image synthesis.
\newblock {\em Advances in Neural Information Processing Systems}, 34:8780--8794, 2021.

\bibitem{dunteman1989principal}
George~H Dunteman.
\newblock {\em Principal components analysis}, volume~69.
\newblock Sage, 1989.

\bibitem{esser2021taming}
Patrick Esser, Robin Rombach, and Bjorn Ommer.
\newblock Taming transformers for high-resolution image synthesis.
\newblock In {\em Proceedings of the 2021 IEEE/CVF Conference on Computer Vision and Pattern Recognition}, pages 12873--12883, 2021.

\bibitem{gatidis2022fdgpetctlesions}
S. Gatidis and T. Kuestner.
\newblock A whole-body fdg-pet/ct dataset with manually annotated tumor lesions (fdg-pet-ct-lesions), 2022.
\newblock Dataset.

\bibitem{goodfellow2014generative}
Ian Goodfellow, Jean Pouget-Abadie, Mehdi Mirza, Bing Xu, David Warde-Farley, Sherjil Ozair, Aaron Courville, and Yoshua Bengio.
\newblock Generative adversarial nets.
\newblock {\em Advances in Neural Information Processing Systems}, 27, 2014.

\bibitem{he2016deep}
Kaiming He, Xiangyu Zhang, Shaoqing Ren, and Jian Sun.
\newblock Deep residual learning for image recognition.
\newblock In {\em Proceedings of the IEEE conference on computer vision and pattern recognition}, pages 770--778, 2016.

\bibitem{isola2017image}
Phillip Isola, Jun-Yan Zhu, Tinghui Zhou, and Alexei~A Efros.
\newblock Image-to-image translation with conditional adversarial networks.
\newblock In {\em Proceedings of the 2017 IEEE Conference on Computer Vision and Pattern Recognition}, pages 1125--1134, 2017.

\bibitem{kingma2014adam}
Diederik~P Kingma and Jimmy Ba.
\newblock Adam: A method for stochastic optimization.
\newblock {\em Computing Research Repository arXiv Preprints arXiv:1412.6980}, 2014.

\bibitem{kong2021breaking}
Lingke Kong, Chenyu Lian, Detian Huang, Yanle Hu, Qichao Zhou, et~al.
\newblock Breaking the dilemma of medical image-to-image translation.
\newblock {\em Advances in Neural Information Processing Systems}, 34:1964--1978, 2021.

\bibitem{lee2019dritdiverseimagetoimagetranslation}
Hsin-Ying Lee, Hung-Yu Tseng, Qi Mao, Jia-Bin Huang, Yu-Ding Lu, Maneesh Singh, and Ming-Hsuan Yang.
\newblock Drit++: Diverse image-to-image translation via disentangled representations, 2019.

\bibitem{li2023bbdm}
Bo Li, Kaitao Xue, Bin Liu, and Yu-Kun Lai.
\newblock {BBDM}: Image-to-image translation with brownian bridge diffusion models.
\newblock In {\em Proceedings of the 2023 IEEE/CVF Conference on Computer Vision and Pattern Recognition}, pages 1952--1961, 2023.

\bibitem{li2020lungpetctdx}
P. Li, S. Wang, T. Li, J. Lu, Y. HuangFu, and D. Wang.
\newblock A large-scale ct and pet/ct dataset for lung cancer diagnosis (lung-pet-ct-dx), 2020.
\newblock Data set.

\bibitem{liang2024leveraging}
Zhiwen Liang, Hui Wei, Gang Liu, Mengjie Cheng, Jiaqi Gao, Song Li, and Xin Tian.
\newblock Leveraging {GAN}-based {CBCT-to-CT} translation models for enhanced image quality and accurate photon and proton dose calculation in adaptive radiotherapy.
\newblock {\em Journal of Radiation Research and Applied Sciences}, 17(1):100809, 2024.

\bibitem{meng2021sdedit}
Chenlin Meng, Yutong He, Yang Song, Jiaming Song, Jiajun Wu, Jun-Yan Zhu, and Stefano Ermon.
\newblock Sdedit: Guided image synthesis and editing with stochastic differential equations.
\newblock {\em arXiv preprint arXiv:2108.01073}, 2021.

\bibitem{milletari2016v}
Fausto Milletari, Nassir Navab, and Seyed-Ahmad Ahmadi.
\newblock V-net: Fully convolutional neural networks for volumetric medical image segmentation.
\newblock In {\em Proceedings of the 4th International Conference on 3D Vision}, pages 565--571, 2016.

\bibitem{muzi2015riderlungpetct}
P. Muzi, M. Wanner, and P. Kinahan.
\newblock Data from rider lung pet-ct, 2015.
\newblock Dataset.

\bibitem{rombach2022high}
Robin Rombach, Andreas Blattmann, Dominik Lorenz, Patrick Esser, and Bj{\"o}rn Ommer.
\newblock High-resolution image synthesis with latent diffusion models.
\newblock In {\em Proceedings of the IEEE/CVF conference on computer vision and pattern recognition}, pages 10684--10695, 2022.

\bibitem{ronneberger2015u}
Olaf Ronneberger, Philipp Fischer, and Thomas Brox.
\newblock {U-Net}: Convolutional networks for biomedical image segmentation.
\newblock In {\em Proceedings of the 18th International Conference on Medical Image Computing and Computer-Assisted Intervention}, pages 234--241, 2015.

\bibitem{saharia2022palette}
Chitwan Saharia, William Chan, Huiwen Chang, Chris Lee, Jonathan Ho, Tim Salimans, David Fleet, and Mohammad Norouzi.
\newblock Palette: Image-to-image diffusion models.
\newblock In {\em Proceedings of the 2022 ACM Special Interest Group on Computer Graphics and Interactive Techniques Conference}, pages 1--10, 2022.

\bibitem{song2020denoising}
Jiaming Song, Chenlin Meng, and Stefano Ermon.
\newblock Denoising diffusion implicit models.
\newblock {\em Computing Research Repository arXiv Preprints arXiv:2010.02502}, 2020.

\bibitem{su2022dual}
Xuan Su, Jiaming Song, Chenlin Meng, and Stefano Ermon.
\newblock Dual diffusion implicit bridges for image-to-image translation.
\newblock {\em Computing Research Repository arXiv Preprints arXiv:2203.08382}, 2022.

\bibitem{tang2019multi}
Hao Tang, Dan Xu, Nicu Sebe, Yanzhi Wang, Jason~J Corso, and Yan Yan.
\newblock Multi-channel attention selection {GAN} with cascaded semantic guidance for cross-view image translation.
\newblock In {\em Proceedings of the 2019 IEEE/CVF Conference on Computer Vision and Pattern Recognition}, pages 2417--2426, 2019.

\bibitem{torbunov2023uvcgan}
Dmitrii Torbunov, Yi Huang, Haiwang Yu, Jin Huang, Shinjae Yoo, Meifeng Lin, Brett Viren, and Yihui Ren.
\newblock Uvcgan: Unet vision transformer cycle-consistent {GAN} for unpaired image-to-image translation.
\newblock In {\em Proceedings of the 2023 IEEE/CVF Winter Conference on Applications of Computer Vision}, pages 702--712, 2023.

\bibitem{upadhyay2021uncertainty}
Uddeshya Upadhyay, Yanbei Chen, Tobias Hepp, Sergios Gatidis, and Zeynep Akata.
\newblock Uncertainty-guided progressive {GANs} for medical image translation.
\newblock In {\em Proceedings of the 24th Internation Conference on Medical Image Computing and Computer Assisted Intervention}, pages 614--624, 2021.

\bibitem{vallieres2017headneckpetct}
Martin Valli{\`e}res, Emily Kay-Rivest, L{\'e}o~Jean Perrin, Xavier Liem, Christophe Furstoss, Nader Khaouam, Phuc~F{\'e}lix Nguyen-Tan, Chang-Shu Wang, and Khalil Sultanem.
\newblock Data from head-neck-pet-ct, 2017.
\newblock Dataset.

\bibitem{vijayakumar2022changing}
Srinivasan Vijayakumar, Johnny Yang, Mary~R Nittala, Alexander~E Velazquez, Brandon~L Huddleston, Nickhil~A Rugnath, Neha Adari, Abhay~K Yajurvedi, Abhinav Komanduri, Claus~Chunli Yang, N~Duggar William, P~Berlin William, Duszak Richard, and Vijayakumar Vani.
\newblock Changing role of {PET/CT} in cancer care with a focus on radiotherapy.
\newblock {\em Cureus}, 14(12), 2022.

\bibitem{wang2023spatial}
Clinton~J Wang, Natalia~S Rost, and Polina Golland.
\newblock Spatial-intensity transforms for medical image-to-image translation.
\newblock {\em IEEE Transactions on Medical Imaging}, 2023.

\bibitem{wang2018highresolutionimagesynthesissemantic}
Ting-Chun Wang, Ming-Yu Liu, Jun-Yan Zhu, Andrew Tao, Jan Kautz, and Bryan Catanzaro.
\newblock High-resolution image synthesis and semantic manipulation with conditional gans, 2018.

\bibitem{wei2021artificial}
Lise Wei and Issam El~Naqa.
\newblock Artificial intelligence for response evaluation with {PET/CT}.
\newblock In {\em Seminars in Nuclear Medicine}, volume~51, pages 157--169, 2021.

\bibitem{yang2020mri}
Qianye Yang, Nannan Li, Zixu Zhao, Xingyu Fan, Eric I-Chao Chang, and Yan Xu.
\newblock Mri cross-modality image-to-image translation.
\newblock {\em Scientific reports}, 10(1):3753, 2020.

\bibitem{zhu2017unpaired}
Jun-Yan Zhu, Taesung Park, Phillip Isola, and Alexei~A Efros.
\newblock Unpaired image-to-image translation using cycle-consistent adversarial networks.
\newblock In {\em Proceedings of the 2017 IEEE International Conference on Computer Vision}, pages 2223--2232, 2017.

\bibitem{zhu2020cross}
Yingying Zhu, Youbao Tang, Yuxing Tang, Daniel~C Elton, Sungwon Lee, Perry~J Pickhardt, and Ronald~M Summers.
\newblock Cross-domain medical image translation by shared latent gaussian mixture model.
\newblock In {\em Medical Image Computing and Computer Assisted Intervention--MICCAI 2020: 23rd International Conference, Lima, Peru, October 4--8, 2020, Proceedings, Part II 23}, pages 379--389. Springer, 2020.

\bibitem{ozbey2023unsupervisedmedicalimagetranslation}
Muzaffer Özbey, Onat Dalmaz, Salman~UH Dar, Hasan~A Bedel, Şaban Özturk, Alper Güngör, and Tolga Çukur.
\newblock Unsupervised medical image translation with adversarial diffusion models, 2023.

\end{thebibliography}
}


\end{document}